\documentclass[english, draftcls, one column]{IEEEtran}
\usepackage[T1]{fontenc}
\usepackage[latin9]{inputenc}
\usepackage{verbatim}
\usepackage{amsmath}
\usepackage{graphicx}
\usepackage{amssymb}
\usepackage{esint}

\makeatletter

\providecommand{\tabularnewline}{\\}

\IEEEoverridecommandlockouts

\usepackage{amsfonts}\usepackage{mathrsfs}\usepackage{bm}\usepackage{algorithmicx}
\usepackage{array}\usepackage{mdwmath}\usepackage{mdwtab}\usepackage[caption=false,font=footnotesize]{subfig}\usepackage{fixltx2e}
\usepackage{footnote}\usepackage{tabularx}\usepackage{multirow}\usepackage{cite}
\usepackage{fixltx2e}\usepackage{braket}\usepackage{algorithmicx}\usepackage{algpseudocode}\usepackage{multirow}
\usepackage{footnote}
\makesavenoteenv{tabular}

\usepackage{url}
\makeatletter

\newcommand{\Rmnum}[1]{\expandafter\@slowromancap\romannumeral #1@}
\makeatother

\makeatother

\usepackage{babel}

\usepackage{url}

\makeatother

\usepackage{babel}

\begin{document}

\title{\textbf{\Large Towards a Theory of Societal Co-Evolution: Individualism
versus Collectivism} }

\author{\IEEEauthorblockN{Kartik Ahuja\IEEEauthorrefmark{1}, Simpson Zhang\IEEEauthorrefmark{2}
and Mihaela van der Schaar\IEEEauthorrefmark{1}}

\author{} \IEEEauthorblockA{Department of Electrical Engineering\IEEEauthorrefmark{1},
Department of Economics\IEEEauthorrefmark{2}, UCLA} }

\maketitle

\thispagestyle{empty} \pagestyle{empty}
\begin{abstract}
Substantial empirical research has shown that the level of individualism
vs. collectivism is one of the most critical and important determinants
of societal traits, such as economic growth, economic institutions
and health conditions. But the exact nature of this impact has thus
far not been well understood in an analytical setting. In this work,
we develop one of the first theoretical models that analytically studies
the impact of individualism-collectivism on the society. We model
the growth of an individual's welfare (wealth, resources and health)
as depending not only on himself, but also on the level of collectivism,
i.e. the level of dependence on the rest of the individuals in the
society, which leads to a co-evolutionary setting. Based on our model,
we are able to predict the impact of individualism-collectivism on
various societal metrics, such as average welfare, average life-time,
 total population, cumulative welfare and average inequality. We analytically
show that individualism has a positive impact on average welfare and
cumulative welfare, but comes with the drawbacks of lower average
life-time, lower total population and higher average inequality. 

%
{}

%
{}

\vspace{-2.5em}

\end{abstract}

\section{Introduction}

Why are some societies wealthier or healthier than others? Why do
some societies have substantial inequality among their members while
others have relatively little? And why do certain societies have a
large population while others have a small population?  Culture, specifically
the level of individualism vs. collectivism in the society, plays
an important and even central role in answering the above questions
\cite{hofstede1980culture}\cite{lenski2005ecological}. 

Landes \cite{landes1998wealth} \cite{landes2000culture} and many
others make the argument for the impact of culture on economic development.
Furthermore, in \cite{gorodnichenko2011dimensions} the authors argue
that among the different dimensions of culture that affect long run
growth, such as individualism-collectivism, masculinity, power distance
etc., the single most relevant dimension  is individualism-collectivism.
Thus understanding the impact of the level of individualism vs. collectivism
on a society is of incredible importance in building a model of societal
development.  In individualistic societies, people tend to depend
more on themselves and less on society for growth in life, whereas
in collectivistic societies, people tend to  contribute to and depend
on society to a greater extent. The level of  collectivism in the
society thus determines how much the growth of an individual is affected
by the society, as well as how much the individual affects the development
of the society, leading to a co-evolutionary setting. In this paper,
collectivism represents a cultural element and not communism or a
state (or religion) direction of activity. 

There has been substantial research \cite{lenski2005ecological}\cite{greif1994cultural}\cite{triandis1985allocentric}\cite{gorodnichenko2010culture}
towards analyzing the determinants of societal development. A significant
thrust of this research has been on developing theories based on empirical
tests\cite{lenski2005ecological}\cite{triandis1985allocentric}\cite{singelis1995horizontal}.
Empirical studies have established the positive impact of individualism
on economic parameters, namely GDP per capita and GDP of a country
\cite{gorodnichenko2010culture}. But there is also more inequality
in the societies with higher levels of development both in economic
\cite{world2012world}\cite{kopczuk2010earnings} and health conditions
\cite{wilkinson1997socioeconomic}. Even though these empirical results
exist, developing mathematical models to understand such social systems
is very important, because these mathematical models help us predict
societal phenomenon and provide useful insights  which can otherwise
not be obtained just based on empirical tests. However, there are
 relatively few papers that analytically study the impact of individualism
vs. collectivism. In \cite{greif1994cultural} the author develops
a mathematical model to show that individualism-collectivism is important
in determining the structure of economic institutions in the society.
In \cite{gorodnichenko2010culture} the authors come up with a mathematical
model through which they can predict that the individualistic societies
promote more long run economic growth  than collectivistic societies.

In this work, we develop a mathematical model of the impact of individualism-collectivism
on more general parameters of a society, as opposed to only on economic
institutions as in the above papers. Our mathematical model helps
us answer questions pertaining to the impact of individualism-collectivism
on the socio-economic inequality in the society, the total population
that can be sustained in the society and the average life-time of
individuals,  which cannot be answered with existing  models. In our
model, individuals are born into the society with a fixed level of
intrinsic quality, which determines the rate of change of their welfare.
Welfare in our model is an abstract quantity which represents an aggregate
of the wealth, resources and health of an individual. An individual
in our model will die either if its level of welfare drops too low
or due to natural causes.  Importantly, the level of collectivism
determines the extent to which an individual's welfare is affected
by rest of the society and vice-versa. Our objective is to compare
societies with different levels of collectivism, levels of welfare
required to survive while assuming the societies are identically impacted
by other factors, such as economic institutions, government \cite{acemoglu2002reversal}\cite{acemoglu2012nations}
or geography, environment \cite{diamond2005guns}. Our model is simplistic
as we abstract away the impact of economic institutions, government,
geography and environment however, it still allows us to capture the
impact of individualism-collectivism, as well as other forces, such
as the level of welfare required to survive on societal metrics, namely
average welfare, average life-time, average inequality, and total
population. From our model we can make the following predictions:

1. Although there is higher societal support given to individuals
with low quality in a collectivistic society, this does not increase
the average welfare of individuals in collectivistic societies since
the support from the rest of individuals in society comes at the expense
of their own welfare levels. This implies lower average welfare levels
in a collectivistic society than in  an individualistic society. 

2. Despite lower average welfare levels,  average life-time may be
higher in a collectivistic society because the social support given
to lower quality individuals will allow them to survive for a longer
amount of time.  This also means that collectivistic societies can
sustain higher population levels. 

3. Cumulative welfare, defined as the total wealth, resources and
health of a society, is lower in a collectivistic society. Although
a collectivistic society supports a larger total population than an
individualistic society, this increase is dominated by the decrease
in the average welfare.

4.  The level of inequality in the society is higher in an individualistic
society than in a collectivistic one, because individualistic societies
allow agents to reach higher  personal welfare while giving less social
support to individuals with low welfare levels.

5. In addition we also study the impact of  rate of birth, rate of
natural deaths and the minimum welfare level required to survive on
the above societal metrics. 

Our  analytical results are in general agreement with the existing
empirical evidence, and we also provide some new predictions that
have so far not been tested empirically. We want to emphasize that
the study here is very general and potentially has a broader scope.
Individualism-collectivism is a trait not particular to humans, and
in a broad sense it can capture the collectivistic versus individualistic
behavior of different biological species, such as bacteria \cite{brown2001cooperation}.
 Being able to mathematically understand individualism and collectivism
is not only useful for societal evolution, but can also be of significant
interest in biology.

%
{}

{}

\vspace{-1em}

\section{System Model}

We consider an infinite horizon continuous-time model with a continuum
of individuals living in a society.  Each individual is characterized
by his intrinsic quality, $Q$, which models his ability to develop
in life, i.e. increase his wealth, resources and health. The intrinsic
quality is a random variable which can take either a good or a bad
value, i.e. $Q\in\{1,-1\},$ where the probability that $Q=1$, $P(Q=1)=\frac{1}{2}$.
Due to space limitations, we only treat a simplistic model here, however
our results  can be extended for more general distributions of quality.
We denote the individual's welfare, an abstract quantity representing
aggregate wealth, resources and health of individual, at time $t$
from birth as, $X(t)$ and the welfare at birth is zero, $X(0)=0$.
The rate at which the welfare of an individual increases at any time
$t$ from birth is determined by the individual's quality as well
as the average quality of  the rest of society, and is given by $R(t)\triangleq\frac{dX(t)}{dt}=(1-w).Q+w.\bar{{\bf Q}}{\bf (t)}$,
where $\bar{{\bf Q}}{\bf (t)}$ is the average quality of all the
individuals in the society at time $t$, and $w\in[0,1]$ is the level
of dependence on society. This weight $w$ is same for all individuals
in the society and is a measure of collectivism in the society, i.e.
$w=1$ and $w=0$ correspond to a purely collectivistic and purely
individualistic society respectively. This mutual dependence amongst
the individuals leads to a co-evolutionary setting.

\begin{figure*}
\centering{}\includegraphics[width=3.2in]{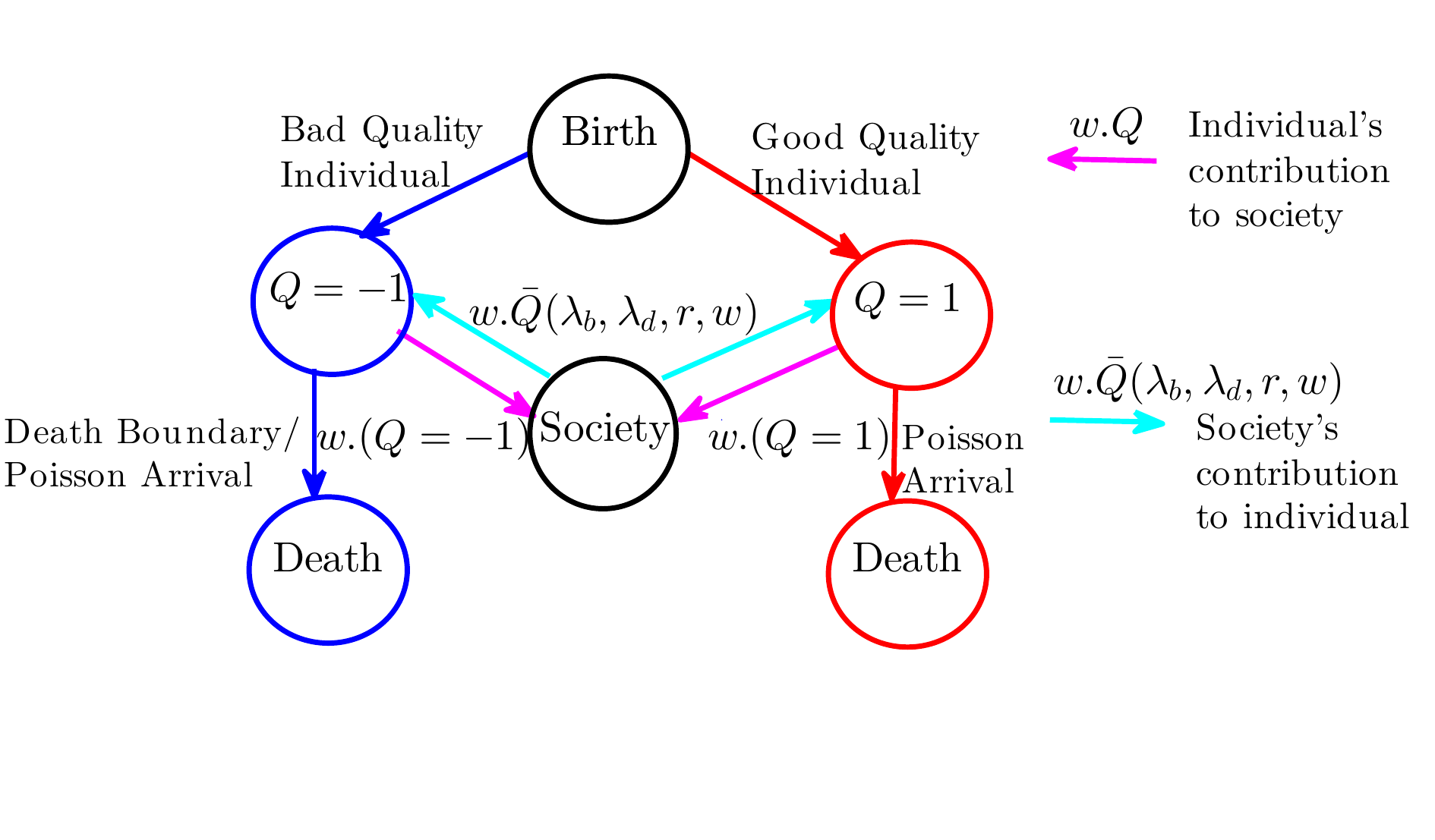}\caption{\label{fig:Life-time-of-good}Life-time of good and bad quality individuals.}
\end{figure*}

The individuals are born into the society at a rate of $\lambda_{b}$
mass per unit time, which means that the total mass of individuals
entering the society in $\Delta t$ time is $\lambda_{b}\Delta t$.
  Individuals in the society can die either due to natural causes
or due to poor welfare levels. The death due to natural causes is
modelled as a Poisson arrival process with a rate $\lambda_{d}$ starting
at the time of birth of the individual, and at the first arrival instance
the individual dies, see Fig. \ref{fig:Life-time-of-good}. The death
due to poor welfare levels happens if the welfare levels fall below
a threshold, $-r$ which we call the death boundary, see Fig. \ref{fig:Life-time-of-good}. 

%
{}

%
{}

\begin{figure*}
\centering{}\includegraphics[width=2.65in]{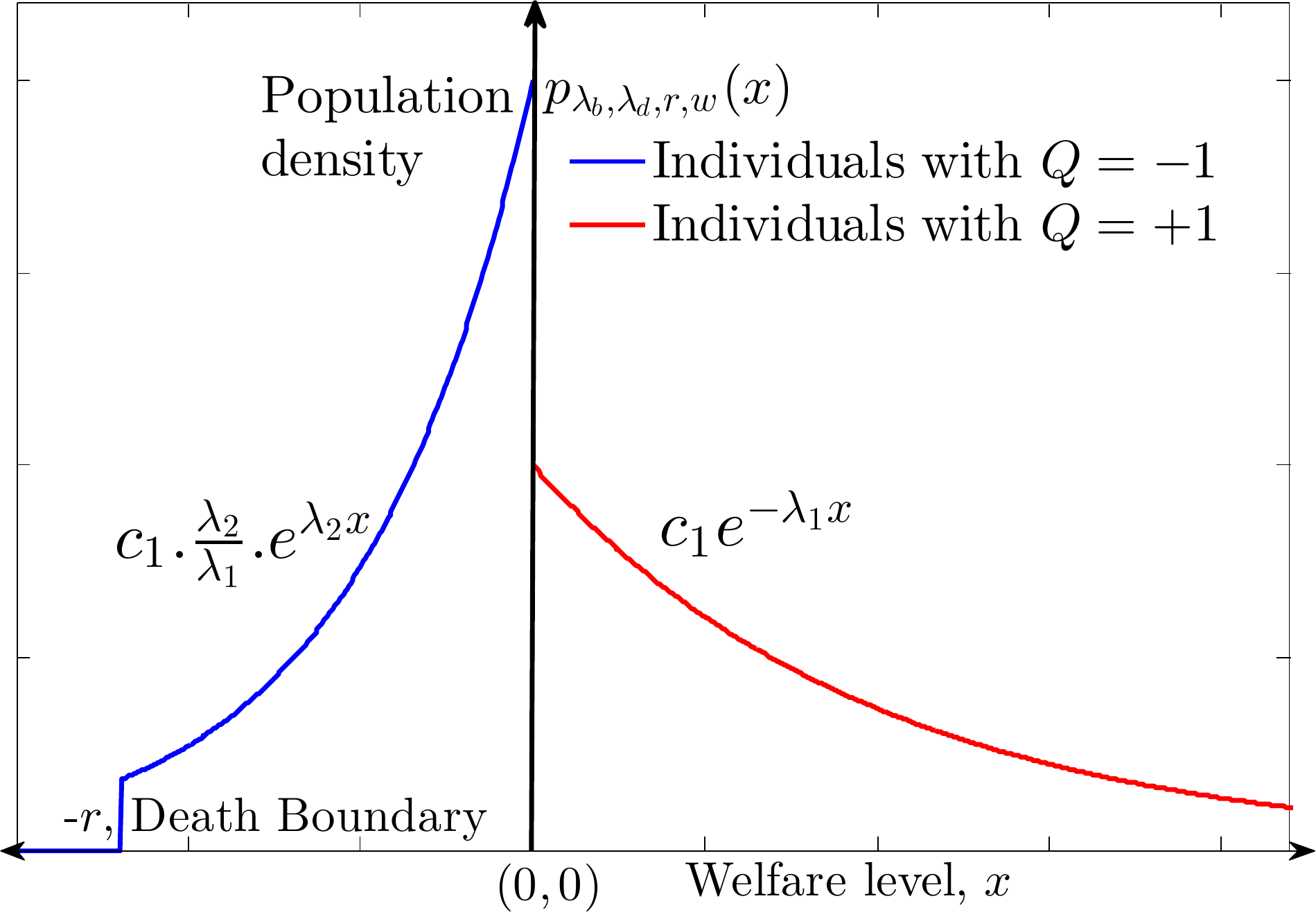}\caption{\label{fig:Steady-State-Distribution}Steady State Distribution of
population density as a function of welfare levels.}
\end{figure*}

$\textit{Steady State of the Society:}$ As time increases the population
increases up to a point where the rate of death equals the rate of
birth. Thus the total population will converge to a fixed mass, and
the distribution of welfare levels within the society will also converge
to a constant. Thus in a steady state: a) the total population mass
in the society attains a fixed value, denoted by $Pop(\lambda_{b},\lambda_{d},r,w)$,
at which the rate of birth will equal the rate of death and b) the
density of the population at a given welfare level $x$, $p_{\lambda_{d},\lambda_{d},r,w}(x)$,
see Fig. \ref{fig:Steady-State-Distribution}, and the mass of the
population with quality $Q=q$, $M(q)$, are also determined. We show
below in Theorem 1 that there is always a unique steady state in our
model given the exogenous parameters $\text{\{}\lambda_{b},\lambda_{d},r,w\text{\}}$,
which characterize the society. 

\textbf{Theorem 1. }Every society has a unique steady state.

 The detailed proofs can be found in the appendix (Section V) given
at the end. 

\textbf{Lemma 1}. Good and bad quality individuals attain positive
and negative welfare values respectively in the steady state. 

Bad quality individuals can die either due to a Poisson arrival or
due to poor welfare levels. As a result the proportion of the bad
quality individuals is lower than that of good quality ones, which
leads to a positive average quality $\bar{Q}(\lambda_{b},\lambda_{d},r,w).$
Hence, good quality individuals cannot take negative welfare values.
Also, it can be shown that the bad quality individuals cannot take
positive welfare values, see the appendix (Section V) at the end for
details. 

In the unique steady state the population density, $p_{\lambda_{d},\lambda_{d},r,w}(x)$
decays exponentially in both positive and negative directions, see
Fig. \ref{fig:Steady-State-Distribution}. We illustrate  the life-time
of an individual with good (bad) quality, i.e. $Q=1\,(Q=-1)$ in steady
state in Fig. \ref{fig:Life-time-of-good}. The positive (negative)
welfare levels are attained by good (bad) quality individuals in the
population.  The rate at which the welfare of a bad quality individual
decays in time is typically lesser than the rate of growth of good
quality individuals, (due to the opposing effects of the negative
quality and positive societal support for a bad quality individual),
this leads to a higher decay in the population density of bad quality
individuals as compared to good quality individuals, see Fig. \ref{fig:Steady-State-Distribution}.
We focus on understanding the impact of the exogenous parameters on
the properties of this steady state. To do so we denominate some important
societal metrics which help understand the properties of the steady
state.

\textbf{Definition 1. Average quality: }The average quality of individuals
represents the net impact the society has on rate of growth of welfare
of each individual and is defined as  $\bar{Q}(\lambda_{b},\lambda_{d},r,w)=1\frac{M(Q=1)}{Pop(\lambda_{b},\lambda_{d},r,w)}-1.\frac{M(Q=-1)}{Pop(\lambda_{b},\lambda_{d},r,w)}$. 

\textbf{Definition 2. Average welfare: }The average value of welfare
of the population, a measure of the average wealth, resources and
health of an individual in the society, is defined as $\bar{X}(\lambda_{b},\lambda_{d},r,w)=\int_{-r}^{\infty}x\frac{p_{\lambda_{b},\lambda_{d},r,w}(x)}{Pop(\lambda_{b},\lambda_{d},r,w)}dx$. 

Let $T$ denote the random variable corresponding to the life-time
of an individual in steady state. Let $R$ be the rate of growth of
the individual in steady state where $R=(1-w).Q+w.\bar{Q}(\lambda_{b},\lambda_{d},r,w))$
and $Q$ is the quality of the individual. If $R\geq0$, then the
individual's welfare will always be above zero, hence the individual
will only die when there is a Poisson arrival. Therefore, $T$ in
this case will be an exponential random variable, $T^{'}$, with mean
$\frac{1}{\lambda_{d}}$. If $R<0$ then the death will happen either
at time $T_{2}(\lambda_{b},\lambda_{d},r,w)=\frac{r}{1-w(1+\bar{Q}(\lambda_{b},\lambda_{d},r,w))}$,
where $T_{2}(\lambda_{b},\lambda_{d},r,w)$ is the time taken to reach
the death boundary, or if there is a Poisson arrival before $T_{2}(\lambda_{b},\lambda_{d},r,w)$.
Hence, $T=\min\{T^{'},T_{2}(\lambda_{b},\lambda_{d},r,w)\}$, 

\textbf{Definition 3. Average life-time: }The average life-time of
an individual is defined as the expected value of the life-time (unconditional
on individual's quality), $\bar{T}(\lambda_{b},\lambda_{d},r,w)=$$E_{\lambda_{b},\lambda_{d},r,w}[T]$. 

%
{}

The next societal metric is a measure of average inequality in the
welfare levels of individuals. 

\textbf{Definition 4. Average inequality: }Average inequality, a measure
of disparity in the society, is defined as  the variance of welfare
, $Var_{X}(\lambda_{b},\lambda_{d},r,w)=\int_{-r}^{\infty}(x-\bar{X}(\lambda_{b},\lambda_{d},r,w))^{2}\frac{p_{\lambda_{b},\lambda_{d},r,w}(x)}{Pop(\lambda_{b},\lambda_{d},r,w)}dx$. 

Next, we come up with a notion of Cumulative welfare, which is the
aggregate amount of welfare in the society, a measure of total wealth
and resources. 

\textbf{Definition 5. Cumulative welfare: }The cumulative welfare,
a measure of total welfare of society accumulated together, is defined
as  $CF(\lambda_{b},\lambda_{d},r,w)=Pop(\lambda_{b},\lambda_{d},r,w)\bar{X}(\lambda_{b},\lambda_{d},r,w)$. 

%
{}

In the above societal metrics, average life-time, total population
and average quality are more intuitive to understand, while average
welfare is similar to GDP per capita \cite{gorodnichenko2010culture},
cumulative welfare is similar to the GDP \cite{gorodnichenko2010culture}
and average inequality is related to GINI coefficient \cite{world2012world}\cite{kopczuk2010earnings}.

\vspace{-1em}

\vspace{-0.35em}

\section{Results}

\vspace{-0.6em}

In this section, we will compare different societal metrics across
societies differing either in the level of collectivism, $w$ or the
other exogenous parameters.  

\textbf{Lemma 2. } a) The average quality $\bar{Q}(\lambda_{b},\lambda_{d},r,w)$
and the average welfare $\bar{X}(\lambda_{b},\lambda_{d},r,w)$ of
an individual decrease as the level of collectivism $w$ increases.
b) $\bar{Q}(\lambda_{b},\lambda_{d},r,w)$ and $\bar{X}(\lambda_{b},\lambda_{d},r,w)$
decrease as the rate of natural deaths $\lambda_{d}$ increases. c)
$\bar{Q}(\lambda_{b},\lambda_{d},r,w)$ and $\bar{X}(\lambda_{b},\lambda_{d},r,w)$
decrease as the  death boundary$-r$ decreases. 

In part a), as the level of collectivism is increased, the support
from the society slows the rate at which the welfare of a bad quality
individual decays with time, causing a larger proportion of the population
to be of low quality. The good quality individuals  contribute more
to this support as well and as a result their own growth is slowed.
As a result, there is a negative impact both on the average quality
and average welfare of the individuals. Parts b) and c) are straightforward,
see the appendix (Section V) at the end for detail. %
{} This lemma is supported by the empirical studies showing lower per
capita income in collectivistic societies in comparison to individualistic
societies\textbf{ }\cite{gorodnichenko2010culture}\textbf{. }

\textbf{}%
{}

%
{}

%
{}

\vspace{-1.3em}

\textbf{Theorem 2.} a) Total population $Pop(\lambda_{b},\lambda_{d},r,w)$
increases as the rate of birth $\lambda_{b}$ increases. b) $Pop(\lambda_{b},\lambda_{d},r,w)$
increases as the level of collectivism $w$ ~increases. c) $Pop(\lambda_{b},\lambda_{d},r,w)$
increases as the death boundary$-r$ decreases. d)  If $w<\frac{1}{2}$
then $Pop(\lambda_{b},\lambda_{d},r,w)$ increases as the rate of
natural deaths $\lambda_{d}$ decreases. %
{}

Part a) and c) are easier to comprehend, see the appendix (Section
V) at the end for details. For part b), as the level of collectivism
increases the support from the society slows the rate at which the
welfare of a bad quality individual decays with time. As a result,
the proportion of individuals dying at the death boundary decreases,
which means that the population level at which the mass of population
dying equals the mass of population being born is higher.\textbf{
}This agrees with the empirical studies which show collectivistic
societies have less income per worker and have a larger population
\cite{triandis1995individualism} \cite{razin1995population}. In
part d), as the rate at which natural deaths occur decreases, the
rate of deaths due to achieving poor welfare levels through hitting
the death boundary can increase. However, if the level of dependence
on the society is low then the decrease in the rate of natural deaths
dominates, and as a result the total population increases such that
the mass of deaths equals mass of birth. 

%
{}

%
{}\textbf{ }

\vspace{-1.3em}

\textbf{Theorem 3:} a) Cumulative welfare $CF(\lambda_{b},\lambda_{d},r,w)$
decreases as the rate of birth $\lambda_{b}$ decreases. b) $CF(\lambda_{b},\lambda_{d},r,w)$
decreases as the rate of natural deaths $\lambda_{d}$ increases.
c) If $\lambda_{d}r\leq\epsilon<\frac{1}{2}\,\,$and $w<\frac{1}{2}-\epsilon$
with $\epsilon>0$, then $CF(\lambda_{b},\lambda_{d},r,w)$ decreases
as the death boundary $-r$ decreases. d) $CF(\lambda_{b},\lambda_{d},r,w)$
decreases as the level of collectivism $w$ increases.

Since $CF(\lambda_{b},\lambda_{d},r,w)\propto Pop(\lambda_{b},\lambda_{d},r,w)$,
part a) follows from Theorem 2. For part b), as the rate of natural
death increases the average welfare of an individual decreases (Lemma
2) and the total population also decreases (Theorem 2), if the level
of collectivism is not high. This shows the result for part b), when
the collectivism is not high. However, it can be shown that even if
the level of collectivism is high then as well there will  be a decrease
in cumulative welfare owing to a significant decrease in the average
welfare (see the appendix (Section V)). For part c), as the death
boundary decreases, the total population in the society increases
whereas the average welfare of an individual decreases, leading to
opposing effects. Therefore, if the $\lambda_{d}r$ is sufficiently
low then the proportion of the population with bad quality is sufficiently
low as well. Also, if the level of collectivism, $w$ is low then
then the rate at which the welfare of bad quality individuals decays
with time is high, hence the effect of decreasing the death boundary
on the average welfare is high. Under these conditions the decrease
in average welfare dominates the increase in population. For part
d), increasing the level of collectivism increases the total population
(Theorem 2), but it decreases the average welfare of an individual
(Lemma 2). Interestingly, it can be shown that the decrease in the
average welfare of an individual dominates the increase in population
(see the appendix (Section V) for technical detail). This result is
also aligned with the empirical tests showing higher GDPs for an individualistic
society \cite{gorodnichenko2010culture}.

\textbf{}%
{}

\textbf{}%
{}

\vspace{-2.3em}

\textbf{Theorem 4.} a) Average life time $\bar{T}(\lambda_{b},\lambda_{d},r,w)$
decreases with an increase in rate of natural deaths $\lambda_{d}$.
b) If $\lambda_{d}r>\theta^{*}=\ln(1+\frac{\sqrt{2}}{2})$ then $\bar{T}(\lambda_{b},\lambda_{d},r,w)$
increases with an increase in level of collectivism $w$ else, it
first decreases and then increases with an increase in level of collectivism
$w$. c), If $\lambda_{d}r>\theta^{*}$, then $\bar{T}(\lambda_{b},\lambda_{d},r,w)$
increases with a decrease in death boundary $-r$ else, it first decreases
and then increases with a decrease in death boundary $-r$.

\begin{figure*}
\centering{}%
\begin{minipage}[t]{0.28\textwidth}%
\includegraphics[width=1.5in]{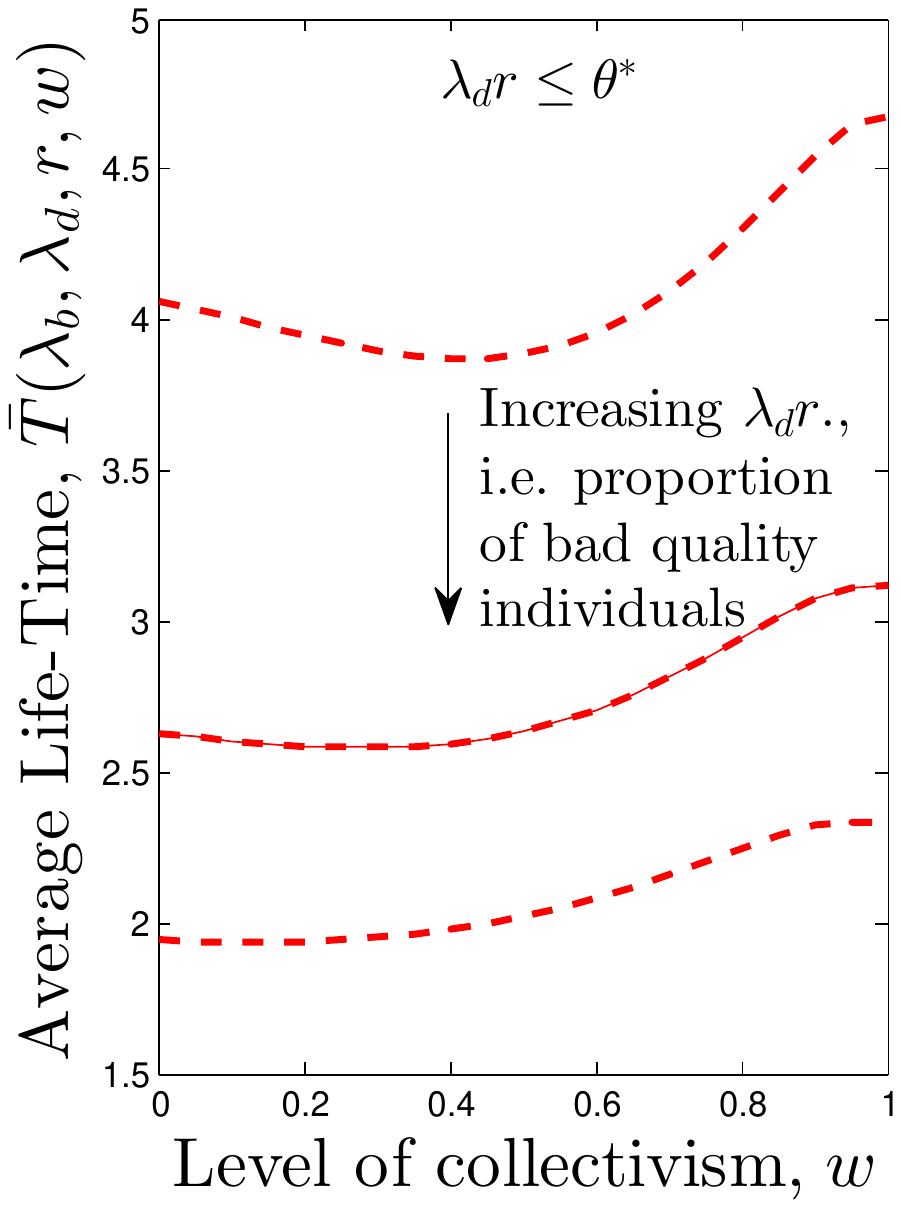}%
\end{minipage}%
\begin{minipage}[t]{0.28\textwidth}%
\includegraphics[width=1.5in]{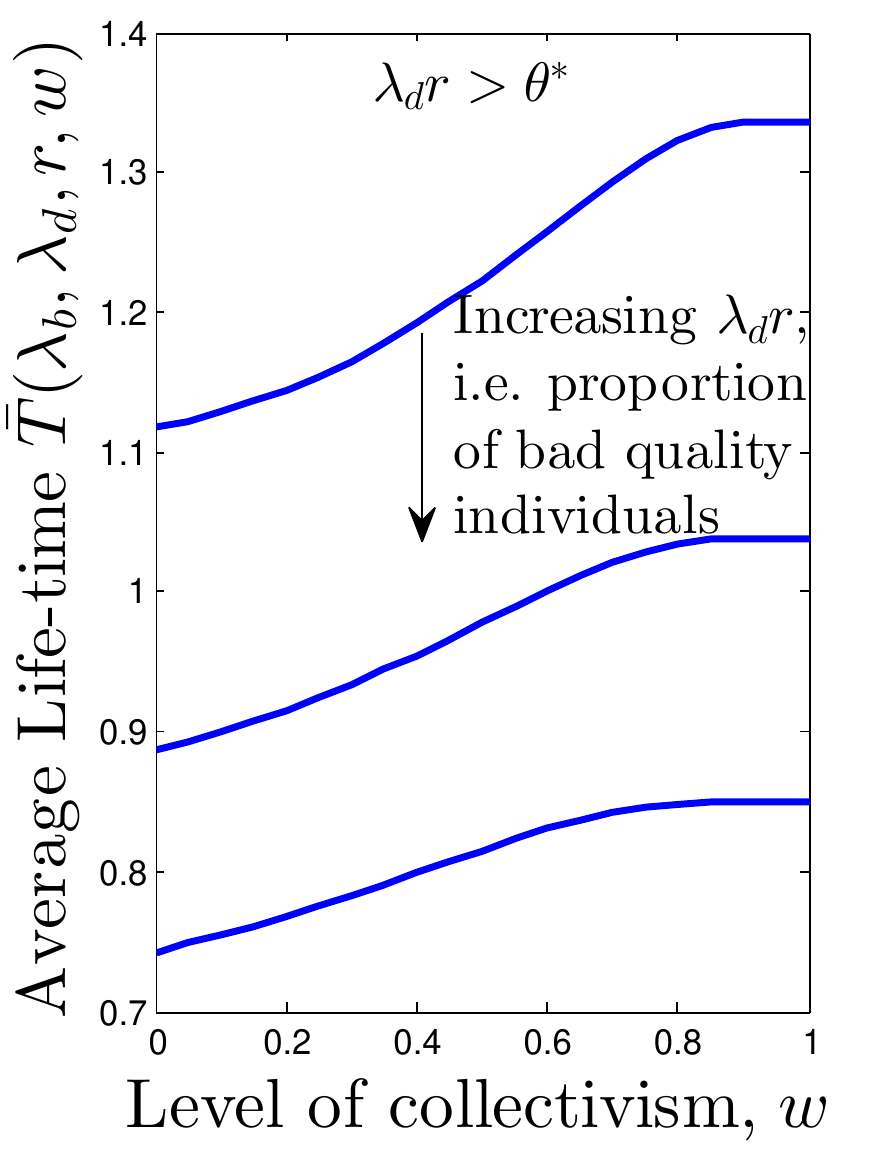}%
\end{minipage}\caption{\label{fig:Illustration-of-part}Illustration of part b) of Theorem
4.}
\end{figure*}

For part a), an increase in the rate of deaths will affect the life-time
of both good and bad quality individuals negatively, thus leading
to the result. For part b), increasing the level of collectivism  slows
the rate at which the welfare of individuals with bad quality decays
with time resulting in an increase in their life-time. It also leads
to an increase in the proportion of individuals with bad quality,
but note that individuals with good quality have a higher life-time
than individuals with bad quality. This leads to an opposing effect.
However, if $\lambda_{d}r$ is high i.e. $\lambda_{d}r>\theta^{*}$,
then the proportion of the individuals with bad quality is high enough,
implying that the increase in the life-time of individuals with bad
quality has a dominating effect in comparison to the decrease resulting
from a decreasing proportion of individuals with good quality. The
proportion of the population of bad quality individuals increases
with an increase in the level of collectivism. If $\lambda_{d}r\leq\theta^{*}$
and the level of collectivism is sufficiently high, there will be
a sufficiently high proportion of bad quality individuals, and so
if level of collectivism is increased then there will be an increase
in the average life-time. However, if the level of collectivism is
not high then there will be a decrease in the average life-time with
an increase in the level of collectivism. A similar explanation applies
to part c). In Fig. \ref{fig:Illustration-of-part}, it is shown that
if $\lambda_{d}r$ is sufficiently high the average life-time increases
with the level of collectivism, otherwise, the average life-time decreases
and then increases. It is important at this point to note that in
part b), we compare two societies with different levels of collectivism
while other parameters remain the same which may include medical facilities,
health awareness etc. that are also crucial determinants of life-time.
Also, our model does not yet consider the impact of cumulative welfare
on the rate of natural deaths $\lambda_{d}$ and is an important direction
for future research.

%
{}

%
{}

%
{}

\vspace{-0.3em}

\textbf{Theorem 5. }The average inequality $Var_{X}(\lambda_{b},\lambda_{d},r,w)$
is always more in an individualistic society $w=0$ as compared to
a collectivistic society $w=1$. Also if the person only dies a natural
death, i.e. $r\rightarrow\infty$, then a) $\lim_{r\rightarrow\infty}Var_{X}(\lambda_{b},\lambda_{d},r,w)$
decreases with an increase in level of collectivism $w$ and b) $\lim_{r\rightarrow\infty}Var_{X}(\lambda_{b},\lambda_{d},r,w)$
decreases with an increase in rate of natural deaths $\lambda_{d}$.

For part a), the case when an individual only dies a natural death
there is a symmetry in the proportion of individuals with good and
bad quality. Hence, the average quality of an individual is zero.
Therefore, the rate of decay (growth) for an individual with bad (good)
quality is $1-w$. Hence, increasing $w$ slows the rate of decay
and growth, thereby allowing individuals to neither take too low or
too high welfare values, which leads to a lower average inequality.
Having higher levels of inequality in individualistic societies has
also been observed in calculations of GINI coefficient for various
countries \cite{world2012world,kopczuk2010earnings}. Also, having
higher inequality has been an important factor affecting the health
of the society \cite{wilkinson1997socioeconomic}, this observation
supports our result on the negative impact of individualism on average
life-time in Theorem 4.\textbf{ }For part b), it is clear that a higher
rate of death, $\lambda_{d}$ implies that individuals with very high
or  low welfare levels are less likely to exist, thus leading to lesser
inequality. 

%
{}%
{}

%
{} 

\vspace{-1.15em}

\section{Conclusion}

\vspace{-.15em} 

We propose a mathematical model to study societal co-evolution under
the forces of individualism and collectivism. This work serves as
an important step towards understanding the exact nature of the impact
of individualism-collectivism on various societal facets. Through
our model we can show that the average welfare of individuals is higher
in an individualistic society, however the average life-time is typically
lower in comparison to a collectivistic society. A larger life-time
in collectivistic society does allow for a larger population to be
sustained, however the cumulative welfare is still lesser. Moreover,
the average inequality is more in an individualistic society owing
to the lack of social support. Our results show concordance with existing
empirical tests.

\section{Appendix}

\textbf{Theorem 1: }Every society has a unique steady state.

\textbf{Proof:} We will start by deriving the population density at
a given welfare level $x$, $p_{\lambda_{b},\lambda_{d},r,w}(x)$
and the total population mass $Pop(\lambda_{d},\lambda_{b},r,w)$
in the steady state and show that they are unique. To do so we first
arrive at the expression for the normalized population density $f_{\lambda_{b},\lambda_{d},r,w}(x)$.
The relation between $f_{\lambda_{b},\lambda_{d},r,w}(x)$, $p_{\lambda_{b},\lambda_{d},r,w}(x)$
and $Pop(\lambda_{d},\lambda_{b},r,w)$ is given as, $Pop(\lambda_{d},\lambda_{b},r,w)=\int_{-\infty}^{\infty}p_{\lambda_{b},\lambda_{d},r,w}(x)dx,\, f_{\lambda_{b},\lambda_{d},r,w}(x)=\frac{p_{\lambda_{b},\lambda_{d},r,w}(x)}{Pop(\lambda_{d},\lambda_{b},r,w)}$.
In steady state the average impact of the society, i.e. $\bar{Q}(\lambda_{b},\lambda_{d},r,w)$
is determined since the proportion of individuals with $Q=q$, i.e.
$M(Q=q)$ do not change. Hence, the rate at which the welfare of an
individual grows can take only two values depending on his quality,
$R_{1}=(1-w).1+w.\bar{Q}(\lambda_{b},\lambda_{d},r,w),\, R_{-1}=\,(1-w).-1+w.\bar{Q}(\lambda_{b},\lambda_{d},r,w),$
here $R_{1}$ and $R_{-1}$ are the rate of growth of good and bad
quality individual respectively. To derive the densities in steady
state, we will first show that in the steady state $R_{1}$ and $R_{-1}$
will be positive and negative respectively. Let's assume that $R_{1}$
and $R_{-1}$ are both positive, i.e. all the individuals in the society
experience a positive growth. In such a case the individuals can only
die due to a Poisson arrival. Also, we know that an individual who
is born is as likely to be good as he is to be bad. Hence, the population
mass at which the rate of death will equal the rate of birth of good/bad
quality individual is the same for both the types of individuals,
i.e. $M(Q=+1)=M(Q=-1)$. As a result, the average quality $\bar{Q}(\lambda_{b},\lambda_{d},r,w)=0$.
Substituting this back in the expressions for the rate we get, $R_{1}=(1-w)$
and $R_{-1}=(1-w).-1$. Therefore, $R_{-1}$ is negative this contradicts
the supposition that the both the rates are positive. Next, let's
assume that both $R_{1}$ and $R_{-1}$ are negative. In this case
the individuals can die either due to a Poisson arrival or due to
hitting the death boundary. In such a case the welfare values attained
will only be negative. Let $f_{\lambda_{b},\lambda_{d},r,w}^{1}(x)$
correspond to the joint density that the individual of good quality
attains a welfare level of $x$. Similarly, we can define $f_{\lambda_{b},\lambda_{d},r,w}^{-1}(x)$
to be the joint density for a bad quality individual at a given welfare
level of $x$. In steady state although the density of population
in a given welfare level is fixed, however the individuals comprising
the density at a given welfare level is not the same owing to change
of welfare levels, births and deaths that happen continually. As a
result, at any instant of time the mass of individuals that attain
a given welfare level will equal the mass of indiduals that leave
that welfare level either due to change in welfare or due to dying.
Consider an infinitesimal interval $h$, the mass of the population
with quality $Q=1$ between $x-h$ and $x$ at time $t$, where $x\leq0$,
is given as, $f_{\lambda_{b},\lambda_{d},r,w}^{1}(x).h$. Consider
a time interval $t^{'}$ after which this mass of individuals, $f_{\lambda_{b},\lambda_{d},r,w}^{1}(x).h$
will either die or will attain a different welfare level between,
$y-h$ and $y$, here $y=x+R_{1}.t^{'}$. The probability that an
individual does not die a natural death in time interval $t^{'}$
is $e^{-\lambda_{d}t^{'}}$. Hence, the proportion of the mass of
individuals who do not die a natural death and a result attain a welfare
between $y-h$ and $y$ is $e^{-\lambda t^{'}}f_{\lambda_{b},\lambda_{d},r,w}^{1}(x).h=f_{\lambda_{b},\lambda_{d},r,w}^{1}(y).h$.
This can be expressed as $f_{\lambda_{b},\lambda_{d},r,w}^{1}(y)=e^{-\lambda_{d}\frac{y-x}{R_{1}}}f_{\lambda_{b},\lambda_{d},r,w}^{1}(x)$
and $f_{\lambda_{b},\lambda_{d},r,w}^{1}(y)=C_{1}.e^{-\lambda_{d}\frac{y}{R_{1}}}$
where $f_{\lambda_{b},\lambda_{d},r,w}^{1}(0)=C_{1}$. Similarly,
for $y\leq0$ we can get $f_{\lambda_{b},\lambda_{d},r,w}^{-1}(y)=C_{-1}.e^{\lambda_{d}\frac{y}{R_{-1}}}$
where $f_{\lambda_{b},\lambda_{d},r,w}^{-1}(0)=C_{-1}$. Note that
both $f_{\lambda_{b},\lambda_{d},r,w}^{1}(x)$ and $f_{\lambda_{b},\lambda_{d},r,w}^{-1}(x)$
are zero for positive welfare values since both good and bad quality
individuals are assumed to have a negative rate of growth. Also, the
rate at which individuals of good quality and bad quality are born
is the same given as $\frac{\lambda_{b}}{2}$. Hence, we can equate
the mass of good (bad) quality individuals which enter the society
in time $\delta t$, i.e. $\frac{\lambda_{b}}{2}\delta t$ to the
mass of individuals between welfare level of $0$ and $\delta x_{1}$
($0$ and $\delta x_{2})$, i.e. $C_{1}\delta x_{1}$ ($C_{-1}\delta x_{-1}$).
This gives, $C_{-1}R_{-1}=C_{1}R_{1}=C$. Since the $f_{\lambda_{b},\lambda_{d},r,w}^{1}(x)$
and $f_{\lambda_{b},\lambda_{d},r,w}^{-1}(x)$ are joint density functions
the integral of the sum of these joint densities should be $1$. \begin{eqnarray*}
 & \int_{-\infty}^{\infty}f_{\lambda_{b},\lambda_{d},r,w}^{1}(x)dx+\int_{-\infty}^{\infty}f_{\lambda_{b},\lambda_{d},r,w}^{-1}(x)dx & =1\\
 & \frac{C_{1}R_{1}}{\lambda_{d}}(1-e^{\frac{\lambda_{d}}{R_{1}}r})+\frac{C_{-1}R_{-1}}{\lambda_{d}}(1-e^{\frac{\lambda_{d}}{R_{-1}}r}) & =1\\
 & C=\frac{\lambda_{d}}{2-e^{\frac{\lambda_{d}}{R_{1}}r}-e^{\frac{\lambda_{d}}{R_{-1}}r}}\end{eqnarray*}

From this we can calculate the mass of the individuals with $Q=1$
and $Q=-1$, i.e. $M(Q=+1)=\frac{C}{\lambda_{d}}(1-e^{\frac{\lambda_{d}}{R_{1}}r})$
and $M(Q=-1)=\frac{C}{\lambda_{d}}(1-e^{\frac{\lambda_{d}}{R_{-1}}r})$.
Since $R_{1}>R_{-1}$ we can see that $M(Q=+1)>M(Q=-1)$. This yields
that the $\bar{Q}(\lambda_{b},\lambda_{d},r,w)>0$ and thereby $R_{1}>0$.
This contradicts the supposition that both the rates are negative.
Also, since $R_{1}>R_{-1}$ the only case left is $R_{1}$ is positive
while $R_{-1}$ is negative. In this case the good and bad quality
individuals take positive and negative welfare values respectively.
We can calculate the joint densities in the same manner as described
above and thus the resulting density is $f_{\lambda_{b},\lambda_{d},r,w}^{1}(x)=C_{1}e^{-\frac{\lambda_{d}}{R_{1}}x},\, x>0$
and $f_{\lambda_{b},\lambda_{d},r,w}^{-1}(x)=C_{-1}e^{\frac{\lambda_{d}}{R-_{1}}x},\, x<0$,
with $C_{1}R_{1}=C_{-1}R_{-1}$. To solve for the constants we need
to proceed in a similar manner as above:

\begin{eqnarray*}
 & \int f_{\lambda_{b},\lambda_{d},r,w}^{1}(x)dx+\int f_{\lambda_{b},\lambda_{d},r,w}^{-1}(x)dx & =1\\
 & \frac{C_{1}R_{1}}{\lambda_{d}}+\frac{C_{-1}R_{-1}}{\lambda_{d}}(1-e^{-\frac{\lambda_{d}}{(1-w).1-w\bar{Q}(\lambda_{b},\lambda_{d},r,w).}r}) & =1\\
 & C=\frac{\lambda_{d}}{2-e^{-\frac{\lambda}{(1-w).1-w\bar{Q}(\lambda_{b},\lambda_{d},r,w).}r})}\end{eqnarray*}

For simplification of notation, we introduce auxiliary notation, $\lambda_{1}=\frac{\lambda_{d}}{(1-w)+w.\bar{Q}(\lambda_{b},\lambda_{d},r,w)}$
and $\lambda_{2}=\frac{\lambda_{d}}{(1-w)-w.\bar{Q}(\lambda_{b},\lambda_{d},r,w)}$.
Hence, the density functions are denoted as follows, $f_{\lambda_{b},\lambda_{d},r,w}^{1}(x)=\frac{\lambda_{1}}{2-e^{-\lambda_{2}r}}e^{-\lambda_{1}x},\, x>0$
and $f_{\lambda_{b},\lambda_{d},r,w}^{-1}(x)=\frac{\lambda_{2}}{2-e^{-\lambda_{2}r}}e^{\lambda_{2}x},\, x<0$.
Also, we can deduce that the marginal density $f_{\lambda_{b},\lambda_{d},r,w}(x)=f_{\lambda_{b},\lambda_{d},r,w}^{1}(x),\, x>0$
and $f_{\lambda_{b},\lambda_{d},r,w}(x)=f_{\lambda_{b},\lambda_{d},r,w}^{-1}(x),\, x<0$.
Using the density computed above we can calculate $M(Q=1)=\frac{1}{2-e^{-\lambda_{2}r}}$
and $M(Q=-1)=\frac{1-e^{-\lambda_{2}r}}{2-e^{-\lambda_{2}r}}$. Also,
the average quality needs to be consistent with the average quality
computed using the distributions derived above. This is formally stated
as \begin{equation}
\bar{Q}(\lambda_{b},\lambda_{d},r,w)=\frac{e^{-\frac{\lambda_{d}r}{1-w-w.\bar{Q}(\lambda_{b},\lambda_{d},r,w)}}}{2-e^{-\frac{\lambda_{d}r}{1-w-w.\bar{Q}(\lambda_{b},\lambda_{d},r,w)}}}=\frac{e^{-\lambda_{2}r}}{2-e^{\lambda_{2}r}}\label{eq:q sol}\end{equation}

Next, we compute the total population mass by equating rate of births
to the rate of deaths. The rate of deaths is comprised of two terms,
the first term is the rate of natural deaths occurring due to Poisson
shocks and the next term is the rate of deaths due to hitting the
death boundary, $f_{\lambda_{b},\lambda_{d},r,w}(-r).(1-w-w.\bar{Q}(\lambda_{b},\lambda_{d},r,w))$
corresponds to the density of the individuals hitting the death boundary
per unit time. Hence, the rate of deaths is $\lambda_{d}.Pop(\lambda_{b},\lambda_{d},r,w)+f_{\lambda_{b},\lambda_{d},r,w}(-r).(1-w-w.\bar{Q}(\lambda_{b},\lambda_{d},r,w)).Pop(\lambda_{b},\lambda_{d},r,w),.$
Equating rate of births to rate of deaths we get the following. 

\[
\lambda_{b}=(\lambda_{d}+f_{\lambda_{b},\lambda_{d},r,w}(-r).(1-w-w.\bar{Q}(\lambda_{b},\lambda_{d},r,w))Pop(\lambda_{b},\lambda_{d},r,w)\]
\[
\lambda_{b}=(\lambda_{d}+\lambda_{d}.\frac{e^{-\lambda_{2}r}}{2-e^{-\lambda_{2}r}}).Pop(\lambda_{b},\lambda_{d},r,w)\]

\begin{equation}
Pop(\lambda_{b},\lambda_{d},r,w)=\frac{\lambda_{b}}{\lambda_{d}(1+\bar{Q}(\lambda_{b},\lambda_{d},r,w))}\end{equation}

Now that we have both the normalized density and the total population's
expressions, we can arrive at the expression of the population density
$p_{\lambda_{b},\lambda_{d},r,w}(x)$ which is just a product of the
two, formally given as follows.

\[
p_{\lambda_{b},\lambda_{d},r,w}(x)=\begin{cases}
Pop(\lambda_{b},\lambda_{d},r,w).\frac{\lambda_{1}}{2-e^{-\lambda_{2}r}}e^{-\lambda_{1}x}, & \text{if }x>0\\
Pop(\lambda_{b},\lambda_{d},r,w).\frac{\lambda_{2}}{2-e^{-\lambda_{2}r}}e^{\lambda_{2}x}, & \mbox{if }x\mbox{<0 }\end{cases}\]

If we can show that there is a unique average quality, $\bar{Q}(\lambda_{b},\lambda_{d},r,w)$
satisfying (1) then both the total population mass (2) and the population
density (3) are uniquely determined. We know that $Q\in\{-1,1\}$
hence, $\bar{Q}(\lambda_{b},\lambda_{d},r,w)\in[-1,1]$. To solve
for $\bar{Q}(\lambda_{b},\lambda_{d},r,w)$, we need to solve $z=g(z)$,
where $g(z)=\frac{e^{-\frac{\lambda_{d}r}{(1-w)-wz.}r}}{2-e^{-\frac{\lambda_{d}r}{1-w-w.z}r}}$
and $z\in[-1,1]$. We will first show that there exists a solution
in the set, $[-1,1]$. Let $z_{1}=-1$ and $z_{2}=\min\{{1,\frac{1-w}{w}}$\}.
If $w<\frac{1}{2}$ then, $z_{2}=1$ else $z_{2}=\frac{1-w}{w}$.
$g(z_{1})=\frac{e^{-\frac{\lambda_{d}r}{1-w}}}{2-e^{-\frac{\lambda_{d}r}{1-w}}}$
and $g(z_{1})>z_{1}$. If $w<\frac{1}{2}$ then $g(z_{2})=\frac{e^{-\frac{\lambda_{d}r}{1-2w}}}{2-e^{-\frac{\lambda_{d}r}{1-2w}}}$
which is less than or equal to $z_{2}=1$, i.e. $g(z_{2})\leq z_{2}$.
Based on this and since the function $z$ and $g(z)$ are continuous
in the range $[-1,\frac{1-w}{w})$, there has to be a point in the
interval $[-1,1]\subset[-1,\frac{1-w}{w})$ where $g(z)=z$. Also,
$g(z)$ is decreasing in the range $[-1,\frac{1-w}{w})$, this can
be seen from the expression for $g^{'}(z)=-\frac{\lambda_{d}rw}{(1-w-wz)^{2}(2e^{\frac{\lambda_{d}r}{1-w-wz}}-1)^{2}}2e^{\frac{\lambda_{d}r}{1-w-wz}}$
and $z$ is strictly increasing function. Therefore, $g(z)-z$ is
a strictly decreasing function in $[-1,\frac{1-w}{w})$, which implies
that the root is unique. When $w=\frac{1}{2}$, $z_{2}=1$ we can
see that $g(z_{1})>z_{1}$ holds, but $g(z)$ is not continuous at
$z_{2}.$ This is not a problem as we know that the function is continuous
everywhere from $[-1,z_{2})$ and $\lim_{z\rightarrow z_{2}}^{'}g(z)=0$,
where $\lim_{z\rightarrow z_{2}}^{'}g(z)$ corresponds to the left
hand limit, hence $\lim_{z\rightarrow z_{2}}^{'}g(z)<z_{2}$. Hence,
the same argument as above can be applied. In the case when $w>\frac{1}{2}$
then we will show that there exists a unique solution for $g(z)=z$
in the range $[-1,1]$. We know that $g(z_{1})=\frac{e^{-\frac{\lambda_{d}r}{1-w}}}{2-e^{-\frac{\lambda_{d}r}{1-w}}}$,
but since $w>\frac{1}{2}$ we need to be careful about the case when
$w=1$. For now we can assume that $\frac{1}{2}<w<1$. Hence, we know
that $g(z_{1})>z_{1}$. Here $z_{2}=\frac{1-w}{w}$ and $g(z)$ will
not be continuous at $z_{2}$. But we can show that $\lim_{z\rightarrow z_{2}}^{'}g(z)=0$,
where $\lim_{z\rightarrow z_{2}}^{'}g(z)$ corresponds to the left
hand limit, and $\lim_{z\rightarrow z_{2}}^{'}g(z)<z_{2}$. Hence,
from the decreasing nature of $g(z)-z$ we know that there is a unique
solution in the range $[-1,\frac{1-w}{w})$. Since $1>w>\frac{1}{2}$
then $[-1,\frac{1-w}{w})\subset[-1,1]$ we need to show that there
is no solution in the range $(\frac{1-w}{w},1]$. In the range $(\frac{1-w}{w},1]$
the function $g(z)$ is not necessarily continuous. There exists a
discontinuity if $2e^{\frac{\lambda_{d}r}{1-w-wz}}-1=0$ and $z\in(\frac{1-w}{w},1]$.
Let's assume that there is a discontinuity. In that case, the function
$g(z)$ will decrease values from $-1$ to $-\infty$, then to the
right of the discontinuity at $2e^{\frac{\lambda_{d}r}{1-w-wz}}-1=0$
the function decreases from $\infty$ to $\frac{1}{2e^{\frac{\lambda_{d}r}{1-2w}}-1}$.
Since $w>\frac{1}{2}$ and $2e^{\frac{\lambda_{d}r}{1-w-wz}}-1=0\:$
for some $z\in(\frac{1-w}{w},1]$ $1>2e^{\frac{\lambda_{d}r}{1-2w}}-1>0$
we can say that $\frac{1}{2e^{\frac{\lambda_{d}r}{1-2w}}-1}>1$. Hence,
there is no point in the range in $[-1,1]$ which intersects with
this function. In the case, when there is no discontinuity it is straightforward
to show that there is no solution of $g(z)=z$ as the function $g(z)$
will only take negative values less than $-1$. Also, when $w=1$
the individuals welfare is fixed to zero all the time, hence there
is a symmetry in the proportion of good and bad quality individuals,
which leads to a unique solution $\bar{Q}(\lambda_{b},\lambda_{d},r,w)=0$. 

\textbf{Lemma 1. }Good and bad quality individuals attain positive
and negative welfare values respectively. 

\textbf{Proof: }The proof of theorem 1, already contains the proof
for this lemma as we show that $R_{1}$ and $R_{-1}$ attain positive
and negative welfare values respectively. 

\textbf{Lemma 2. }The average quality $\bar{Q}(\lambda_{b},\lambda_{d},r,w)$
and the average welfare $\bar{X}(\lambda_{b},\lambda_{d},r,w)$ of
an individual a). Decrease as the level of collectivism, $w$ is increased.,
b). Decrease as the rate of natural deaths, $\lambda_{d}$ increases.,
c). Decrease as the the death boundary,$-r$ decreases. 

\textbf{Proof: }We already know that the solution for $\bar{Q}(\lambda_{b},\lambda_{d},r,w)$
requires solving a transcendental equation (1), which means that we
do not have a closed form analytical expression for it. It can be
shown that the expression for $\bar{X}(\lambda_{b},\lambda_{d},r,w)$
expressed in terms of the $\bar{Q}(\lambda_{b},\lambda_{d},r,w)$
is $(r+\frac{1}{\lambda_{d}}).\bar{Q}(\lambda_{b},\lambda_{d},r,w)$.
From Theorem 1, we know that for every set of parameters there does
exist a solution $\bar{Q}(\lambda_{b},\lambda_{d},r,w)$. For part
a), as the level of collectivism is increased let us assume that the
average quality $\bar{Q}(\lambda_{b},\lambda_{d},r,w)$ increases.
However, if there is an increase in both the collectivism and the
average quality, the expression $g(\bar{Q}(\lambda_{b},\lambda_{d},r,w))=\frac{e^{-\frac{\lambda_{d}r}{1-w-w.\bar{Q}(\lambda_{b},\lambda_{d},r,w)}}}{2-e^{-\frac{\lambda_{d}r}{1-w-w.\bar{Q}(\lambda_{b},\lambda_{d},r,w)}}}$
decreases which contradicts the increase in $\bar{Q}(\lambda_{b},\lambda_{d},r,w)$.
Hence, $\bar{Q}(\lambda_{b},\lambda_{d},r,w)$ has to decrease with
an increase in collectivism. And from the expression of $\bar{X}(\lambda_{b},\lambda_{d},r,w)$
expressed in terms of $\bar{Q}(\lambda_{b},\lambda_{d},r,w)$ it is
straightforward that the average welfare also decreases with an increase
in the level of collectivism. For part b), again as the rate of natural
deaths increases assume that $\bar{Q}(\lambda_{b},\lambda_{d},r,w)$
increases. However the decrease in $\frac{e^{-\frac{\lambda_{d}r}{1-w-w.\bar{Q}(\lambda_{b},\lambda_{d},r,w)}}}{2-e^{-\frac{\lambda_{d}r}{1-w-w.\bar{Q}(\lambda_{b},\lambda_{d},r,w)}}}$
will contradict the assumption. With an increase in $\lambda_{d}$
the first term in the expression of $\bar{X}(\lambda_{b},\lambda_{d},r,w)$
which inversely related to $\lambda_{d}$ has to decrease, this combined
with the decrease in $\bar{Q}(\lambda_{b},\lambda_{d},r,w)$ leads
to a decrease in the average welfare. For part c), we arrive at the
expression of the derivative of average quality $\bar{Q}(\lambda_{b},\lambda_{d},r,w)$
w.r.t. $r$, $-\frac{\lambda_{d}(d)(d+1)(1-w-wd)}{(1-w-wd)^{2}+\lambda_{d}rw(d+1)}$
which is negative. Hence, we know that the average quality indeed
decreases with an increase in $r$. For average welfare we give an
intuitive explanation first, increasing $r$ decreases the average
quality as a result of which the growth of a good quality individual
slows down and the decay of a bad quality individual becomes faster.
As a result the average welfare levels attained by a good and bad
quality individual are lower. Moreover, increase in $r$ increases
the proportion of the bad quality individuals which further has a
negative effect on the average welfare. To prove this formally we
will show that the average welfare of both good and bad quality individuals
decreases and the proportion of the bad quality individuals increases.
Since the average welfare value of a bad quality individual is always
lower than that of a good quality individual this is sufficient to
show the result. The average welfare of good quality individuals is
given as $\frac{1}{\lambda_{1}}=\frac{1-w+w\bar{Q}(\lambda_{b},\lambda_{d},r,w)}{\lambda_{d}}$.
This can be derived as follows, the distribution of the welfare conditional
on the fact that individuals are of good quality $f_{\lambda_{b},\lambda_{d},r,w}(x|Q=+1)$
can be shown to be an exponential distribution with parameter $\lambda_{1}$
exactly on the same lines as we derived the joint densities $f_{\lambda_{b},\lambda_{d},r,w}^{1}(x)$
in Theorem 1. Since $\bar{Q}(\lambda_{b},\lambda_{d},r,w)$ decreases
as a function of $r$, the average welfare of a good quality individual
also decreases as a function of $r$. Similarly we need to arrive
at the distribution $f_{\lambda_{b},\lambda_{d},r,w}(x|Q=-1),$which
turns out to be $f_{\lambda_{b},\lambda_{d},r,w}(x|Q=-1)=\frac{\lambda_{2}}{1-e^{-\lambda_{2}r}}e^{\lambda_{2}x},\, x<0$.
The average welfare value of bad quality individual can be arrived
at using this distribution and it turns out to be, $-\frac{1}{\lambda_{2}}+\frac{re^{-\lambda_{2}r}}{1-e^{-\lambda_{2}r}}$.
As $r$ is increased, $\bar{Q}(\lambda_{b},\lambda_{d},r,w)$ decreases
and thus $\lambda_{2}$ decreases as well. The partial derivative
of average welfare of bad quality individual $-\frac{1}{\lambda_{2}}+\frac{re^{-\lambda_{2}r}}{1-e^{-\lambda_{2}r}}$
w.r.t. $r$ is given as $\frac{e^{\lambda_{2}r}-\lambda_{2}re^{\lambda_{2}r}-1}{(e^{\lambda_{2}r}-1)^{2}}$
and this expression turns out to be negative for $(\lambda_{2},r)\in\mathbb{R}_{+}^{2}$.
Also, it can be shown that the partial derivative of $-\frac{1}{\lambda_{2}}+\frac{re^{-\lambda_{2}r}}{1-e^{-\lambda_{2}r}}$
w.r.t $\lambda_{2}$ is given as $(\frac{1}{\lambda_{2}})^{2}-\frac{r^{2}e^{\lambda_{2}r}}{(e^{\lambda_{2}r}-1)^{2}}$
and this expression turns out to be positive. Hence, from the sign
of these partial derivatives we can easily see the result. 

\vspace{-0.6em}

\textbf{Theorem 2. }a) Total population $Pop(\lambda_{b},\lambda_{d},r,w)$
increases as the rate of birth $\lambda_{b}$ increases. b) $Pop(\lambda_{b},\lambda_{d},r,w)$
increases as the level of collectivism $w$ ~increases. c) $Pop(\lambda_{b},\lambda_{d},r,w)$
increases as the death boundary$-r$ decreases. d)  If $w<\frac{1}{2}$
then $Pop(\lambda_{b},\lambda_{d},r,w)$ increases as the rate of
natural deaths $\lambda_{d}$ decreases. %
{} %
{}

\textbf{Proof: }In order to compute the total population in the steady
state, we need to have the rate of birth equals the rate of death
which is formally stated as follows,

\[
(\lambda_{d}+f_{\lambda_{b},\lambda_{d},r,w}(-r).(1-w-w.\bar{Q}(\lambda_{b},\lambda_{d},r,w))).Pop(\lambda_{b},\lambda_{d},r,w)=\lambda_{b}\]

\[
Pop(\lambda_{b},\lambda_{d},r,w)=\frac{\lambda_{b}}{\lambda_{d}.(1+\bar{Q}(\lambda_{b},\lambda_{d},r,w))}\]

For part a), $\bar{Q}(\lambda_{b},\lambda_{d},r,w)$ does not depend
on the rate of births and it is clear that the result holds since
the population is directly proportional to $\lambda_{b}$. For part
b) as well it can be seen that the only term in the expression which
depends on $w$ is $\bar{Q}(\lambda_{b},\lambda_{d},r,w)$ which will
decrease as $w$ is increased (Lemma 2). Therefore, it is clear that
the population has to increase with level of collectivism. For part
c), again we can see that the only term in the expression which depends
on the death boundary $-r$ is $\bar{Q}(\lambda_{b},\lambda_{d},r,w)$.
We know that as the death boundary decreases $\bar{Q}(\lambda_{b},\lambda_{d},r,w)$
decreases as well (Lemma 2), thereby leading to an increase in the
population. In part d), as the rate at which natural deaths occur
decreases, the rate of deaths due to achieving poor welfare levels
or hitting the death boundary can increase. However, if the level
of dependence on the society is low then the decrease in the rate
of natural deaths dominates, as a result the total population increases
such that the mass of deaths equals mass of birth. We now show this
formally. Let us take the derivative of the term in the denominator
w.r.t $\lambda_{d}$,

\[
(1+\bar{Q}(\lambda_{b},\lambda_{d},r,w))+\frac{d\bar{Q}(\lambda_{b},\lambda_{d},r,w)}{d\lambda_{d}}\]

\[
(1+\bar{Q}(\lambda_{b},\lambda_{d},r,w))(\frac{(1-w-w.\bar{Q}(\lambda_{b},\lambda_{d},r,w))^{2}+\lambda_{d}rw\bar{Q}(\lambda_{b},\lambda_{d},r,w)-\lambda_{d}r\bar{Q}(\lambda_{b},\lambda_{d},r,w)(1-w-w.\bar{Q}(\lambda_{b},\lambda_{d},r,w))}{(1-w-w.\bar{Q}(\lambda_{b},\lambda_{d},r,w))^{2}+\lambda_{d}rw\bar{Q}(\lambda_{b},\lambda_{d},r,w)}\]

\[
(1+\bar{Q}(\lambda_{b},\lambda_{d},r,w))(\frac{(1-w-w.\bar{Q}(\lambda_{b},\lambda_{d},r,w))(1-w-w.\bar{Q}(\lambda_{b},\lambda_{d},r,w)-\lambda_{d}r\bar{Q}(\lambda_{b},\lambda_{d},r,w))+\lambda_{d}rw\bar{Q}(\lambda_{b},\lambda_{d},r,w)}{(1-w-w.\bar{Q}(\lambda_{b},\lambda_{d},r,w))^{2}+\lambda_{d}rw\bar{Q}(\lambda_{b},\lambda_{d},r,w)})\]

If we can show that $(1-w-w.\bar{Q}(\lambda_{b},\lambda_{d},r,w)-\lambda_{d}r\bar{Q}(\lambda_{b},\lambda_{d},r,w))>0$
then the above expression will be positive. We know from lemma 2 that
$\bar{Q}(\lambda_{b},\lambda_{d},r,0)\geq\bar{Q}(\lambda_{b},\lambda_{d},r,w),\,\forall w\in[0,1]$
. This leads to $\bar{Q}(\lambda_{b},\lambda_{d},r,0)<\frac{1-w}{w+\lambda_{d}r}$
which is a sufficient for the above derivative to be positive. It
can be checked that this condition is satisfied if $w<\frac{1}{2}$. 

%
{}

%
{}\textbf{ }

\vspace{-0.8em}

\textbf{Theorem 3: }a) Cumulative welfare $CF(\lambda_{b},\lambda_{d},r,w)$
decreases as the rate of birth $\lambda_{b}$ decreases. b) $CF(\lambda_{b},\lambda_{d},r,w)$
decreases as the rate of natural deaths $\lambda_{d}$ increases.
c) If $\lambda_{d}r\leq\epsilon<\frac{1}{2}\,\&\, w<\frac{1}{2}-\epsilon$~with
$\epsilon>0$, then $CF(\lambda_{b},\lambda_{d},r,w)$ decreases as
the death boundary $-r$ decreases. d) $CF(\lambda_{b},\lambda_{d},r,w)$
decreases as the level of collectivism $w$ increases.

\textbf{Proof: }For part a), we know that $CF(\lambda_{b},\lambda_{d},r,w)=\bar{X}(\lambda_{b},\lambda_{d},r,w)Pop(\lambda_{b},\lambda_{d},r,w)$.
Also, since the average welfare of an individual is independent of
$\lambda_{b}$ we only need to consider the effect on total population
which we already know from Theorem 2. For part b), let us simplify
the expression of cumulative welfare, $CF(\lambda_{b},\lambda_{d},r,w)=(r+\frac{1}{\lambda_{d}}).\frac{\lambda_{b}}{\lambda_{d}}.\frac{\bar{Q}(\lambda_{b},\lambda_{d},r,w)}{(1+\bar{Q}(\lambda_{b},\lambda_{d},r,w))}$
. From this expression we can see that as $\lambda_{d}$ increases
the term $(r+\frac{1}{\lambda_{d}}).\frac{\lambda_{b}}{\lambda_{d}}$
will definitely decrease. In fact the other term will also decrease,
as can be seen from the derivative of the second term w.r.t. $\lambda_{d}$
, $\frac{1}{(1+\bar{Q}(\lambda_{b},\lambda_{d},r,w))^{2}}\frac{d\bar{Q}(\lambda_{b},\lambda_{d},r,w)}{d\lambda_{d}}$
and this combined with Lemma 2. For part d), we can see that only
$\frac{\bar{Q}(\lambda_{b},\lambda_{d},r,w)}{(1+\bar{Q}(\lambda_{b},\lambda_{d},r,w))}$
depends on the weight $w$ and its derivative w.r.t. $w$ is $\frac{1}{(1+\bar{Q}(\lambda_{b},\lambda_{d},r,w))^{2}}\frac{d\bar{Q}(\lambda_{b},\lambda_{d},r,w)}{dw}$.
This expression of the derivative and Lemma 2, lead us to the result.
For part c), as the death boundary decreases, the total population
in the society increases whereas the average welfare of an individual
decreases, leading to opposing effects. Therefore, if the $\lambda_{d}r$
is sufficiently low then the proportion of the population with bad
quality is sufficiently low as well. Also, if the level of collectivism,
$w$ is low then then the rate at which the welfare of bad quality
individuals decays with time is high, hence the effect of decreasing
the death boundary on the average welfare is high. Under these conditions
the decrease in average welfare dominates the increase in population.
We next show this formally. The derivative of cumulative welfare w.r.t.
$r$ is given as, 

\[
\big(\frac{\bar{Q}(\lambda_{b},\lambda_{d},r,w)}{\bar{Q}(\lambda_{b},\lambda_{d},r,w)+1}\big).(\frac{(1-w-w.\bar{Q}(\lambda_{b},\lambda_{d},r,w)^{2}-(\lambda_{d}r+1)(1-w-w.\bar{Q}(\lambda_{b},\lambda_{d},r,w)))}{(1-w-w.\bar{Q}(\lambda_{b},\lambda_{d},r,w)^{2}+\lambda_{d}rwd(d+1)})\]

\[
\big(\frac{\bar{Q}(\lambda_{b},\lambda_{d},r,w)}{\bar{Q}(\lambda_{b},\lambda_{d},r,w)+1}\big).(\frac{(w.(\bar{Q}(\lambda_{b},\lambda_{d},r,w)).(-(1-w.(1+\bar{Q}(\lambda_{b},\lambda_{d},r,w))+\lambda_{d}r+\lambda_{d}r\bar{Q}(\lambda_{b},\lambda_{d},r,w))-\lambda_{d}r}{(1-w-w.\bar{Q}(\lambda_{b},\lambda_{d},r,w)^{2}+\lambda_{d}rwd(d+1)})\]

If $-(1-w.(1+\bar{Q}(\lambda_{b},\lambda_{d},r,w))+\lambda_{d}r+\lambda_{d}r\bar{Q}(\lambda_{b},\lambda_{d},r,w)<0$
then the above derivative is negative. Note $\bar{Q}(\lambda_{b},\lambda_{d},r,0)<\frac{1-w-\lambda_{d}r}{w+\lambda_{d}r}$
is sufficient for this condition to hold and it leads to the following
condition, $w<\frac{1}{2}-\epsilon$ and $\lambda_{d}r\leq\epsilon<\frac{1}{2}$
where $\epsilon>0$. This proves part c. 

\textbf{}%
{}

\textbf{}%
{}

\vspace{-1.4em}

\textbf{Theorem 4.} a) Average life time $\bar{T}(\lambda_{b},\lambda_{d},r,w)$
decreases with an increase in rate of natural deaths $\lambda_{d}$.
b) If $\lambda_{d}r>\theta^{*}=\ln(1+\frac{\sqrt{2}}{2})$%
\footnote{$\theta^{*}$ is a fixed constant which in general will depend on
$P(Q=1),$ and when $P(Q=1)=\frac{1}{2}$ it is $\ln(1+\frac{\sqrt{2}}{2}).$%
}, then $\bar{T}(\lambda_{b},\lambda_{d},r,w)$ increases with an increase
in level of collectivism $w$ else, it first decreases and then increases
with an increase in level of collectivism $w$. c), If $\lambda_{d}r>\theta^{*}$,
then $\bar{T}(\lambda_{b},\lambda_{d},r,w)$ increases with a decrease
in death boundary $-r$ else, it first decreases and then increases
with a decrease in death boundary $-r$. 

\textbf{Proof: }The expression for the average life-time of an individual
$\bar{T}(\lambda_{b},\lambda_{d},r,w)$ involves the computation of
the average life-time of good quality individuals and bad quality
individuals separately and then combining the two using the conditional
probabilities. Hence, $\bar{T}(\lambda_{b},\lambda_{d},r,w)=\frac{1}{\lambda_{d}}+(\frac{1}{\lambda_{d}})\frac{\bar{Q}(\lambda_{b},\lambda_{d},r,w)-\bar{Q}(\lambda_{b},\lambda_{d},r,w)^{2}}{\bar{Q}(\lambda_{b},\lambda_{d},r,w)+1}$.
The derivative of $\bar{T}(\lambda_{b},\lambda_{d},r,w)$ w.r.t $w$
can be expressed as $\frac{1}{\lambda_{d}}\frac{\bar{Q}(\lambda_{b},\lambda_{d},r,w)^{2}+2\bar{Q}(\lambda_{b},\lambda_{d},r,w)-1}{\bar{(Q}(\lambda_{b},\lambda_{d},r,w)+1)^{2}}\frac{d\bar{Q}(\lambda_{b},\lambda_{d},r,w)}{dw}$.
If $\bar{Q}(\lambda_{b},\lambda_{d},r,0)<\sqrt{2}-1$ then the above
derivative is positive. This leads to the condition $\lambda_{d}r>\ln(1+\frac{\sqrt{2}}{2})$.
However, if $\lambda_{d}r<\ln(1+\frac{\sqrt{2}}{2})$ then $\bar{Q}(\lambda_{b},\lambda_{d},r,0)>\sqrt{2}-1$
and as a result the derivative is negative. However, $\bar{Q}(\lambda_{b},\lambda_{d},r,w)$
will decrease with increase in $w$ and it can be observed that at
$w=1$, $\bar{Q}(\lambda_{b},\lambda_{d},r,w)$ will be zero, this
is due to the fact that the individuals completely depend on the society
and the rate of growth is zero for all individuals. Hence, for some
$w=w^{*}$ the $\bar{Q}(\lambda_{b},\lambda_{d},r,w^{*})=\sqrt{2}-1$
where the life-time will take the minimum value. Therefore, we know
that in the region $w>w^{*}$, the life-time will increase. This explains
part b). For part c), a similar explanation can be given. The expression
for the derivative changes to $\frac{1}{\lambda_{d}}\frac{\bar{Q}(\lambda_{b},\lambda_{d},r,w)^{2}+2\bar{Q}(\lambda_{b},\lambda_{d},r,w)-1}{\bar{(Q}(\lambda_{b},\lambda_{d},r,w)+1)^{2}}\frac{d\bar{Q}(\lambda_{b},\lambda_{d},r,w)}{dr}$
and the rest of the explanation follows from above and Lemma 2. For
part a), we will first show that the average life-time of both a good
and bad quality individual decrease. Then, we will show that the proportion
of the bad quality individuals increase. Since the average life-time
of a bad quality individual is always lesser than that of a good quality
individual, this will lead to a decrease in the average life-time
unconditional on the quality of the individual. First of all the average
life-time of a good quality individual is $\frac{1}{\lambda_{d}}$
and it decreases with $\lambda_{d}$. Next, the average life-time
of an individual with bad quality is arrived at by computing the expectation
of $\min\{T^{'},T_{2}(\lambda_{b},\lambda_{d},r,w)=\frac{r}{1-w(1+\bar{Q}(\lambda_{b},\lambda_{d},r,w))},\}$
where $T^{'}$ is an exponential random variable with mean $\frac{1}{\lambda_{d}}$.
The life-time of a bad quality individual is $\frac{1}{\lambda_{d}}.(1-e^{-\frac{\lambda_{d}r}{1-w-w.\bar{Q}(\lambda_{b},\lambda_{d},r,w)}})$,
the derivative of this expression is $-\frac{1}{\lambda_{d}^{2}}.(1-e^{-\frac{\lambda_{d}r}{1-w-w.\bar{Q}(\lambda_{b},\lambda_{d},r,w)}}-\frac{\lambda_{d}r}{1-w-w\bar{Q}(\lambda_{b},\lambda_{d},r,w)}e^{-\frac{\lambda_{d}r}{1-w-w.\bar{Q}(\lambda_{b},\lambda_{d},r,w)}})+\frac{r.w}{1-w-w.\bar{Q}(\lambda_{b},\lambda_{d},r,w)}\frac{d\bar{Q}(\lambda_{b},\lambda_{d},r,w)}{d\lambda_{d}}$.
The term $(1-e^{-\frac{\lambda_{d}r}{1-w-w.\bar{Q}(\lambda_{b},\lambda_{d},r,w)}}-\frac{\lambda_{d}r}{1-w-w\bar{Q}(\lambda_{b},\lambda_{d},r,w)}e^{-\frac{\lambda_{d}r}{1-w-w.\bar{Q}(\lambda_{b},\lambda_{d},r,w)}})$
has to be positive since $(x+1)e^{-x}<1$. Hence, we can see that
the derivative is negative which implies the result. 

%
{}

%
{}

%
{}

\textbf{Theorem 5. }The average inequality $Var_{X}(\lambda_{b},\lambda_{d},r,w)$
is always more in an individualistic society $w=0$ as compared to
a collectivistic society $w=1$. Also if the person only dies a natural
death, i.e. $r\rightarrow\infty$, then a) $\lim_{r\rightarrow\infty}Var_{X}(\lambda_{b},\lambda_{d},r,w)$
decreases with an increase in level of collectivism $w$ and b) $\lim_{r\rightarrow\infty}Var_{X}(\lambda_{b},\lambda_{d},r,w)$
decreases with an increase in rate of natural deaths $\lambda_{d}$.

\textbf{Proof:~}$Var_{X}(\lambda_{b},\lambda_{d},r,w=1)=0$ since
all the individuals have the same welfare value of zero. So, we need
to show that $Var_{X}(\lambda_{b},\lambda_{d},r,w=0)>0$. The expression
for variance is, \[
Var_{X}(\lambda_{b},\lambda_{d},r,w=0)=(\frac{1}{\lambda_{d}})^{2}\frac{(8e^{2\lambda_{d}r}+e^{\lambda_{d}r}(-2(\lambda_{d}r)^{2}+4\lambda_{d}r-8)-3\lambda_{d}r+(\lambda_{d}r)^{2}+1)}{(2e^{\lambda_{d}r}-1)^{2}}\]

It can be shown that the expression in the numerator of the above
expression is indeed positive. To do so we show that at any point
$(\lambda_{d},r)\in\mathbb{R}_{+}^{2}$the partial derivative w.r.t
to either $\lambda_{d}$ or $r$ is positive and also that $Var_{X}(\lambda_{b},\lambda_{d}=0,r=0,w=0)>0$
which helps us establish the result. 

For part a), the case when an individual only dies a natural death
there is a symmetry in the proportion of individuals with good and
bad quality. Hence, the average quality of an individual is zero.
Therefore, the rate of decay for an individual with bad quality is
$1-w$ and the same is the rate of growth for an individual with good
quality. Hence, increasing $w$ slows the rate of decay and growth,
thereby allowing individuals to neither take too low or too high welfare
values, which leads to a lower average disparity. Formally if $r\rightarrow\infty$,
$\bar{Q}(\lambda_{b},\lambda_{d},r,w)=0$, this leads to the density
distribution given as, $f_{\lambda_{b},\lambda_{d},r,w}^{1}(x)=\frac{\lambda_{d}}{1-w}e^{-\frac{\lambda_{d}}{1-w}x}$
and $f_{\lambda_{b},\lambda_{d},r,w}^{-1}(x)=\frac{\lambda_{d}}{1-w}e^{\frac{\lambda_{d}}{1-w}x}$.
This leads to the expression of the variance given as, $(\frac{1-w}{\lambda_{d}})^{2}$
and therefore, part a) and b) follow directly from this. 

{}

\bibliographystyle{IEEEtran}
\bibliography{SIP_ref}

\begin{thebibliography}{10}
\providecommand{\url}[1]{#1}
\csname url@samestyle\endcsname
\providecommand{\newblock}{\relax}
\providecommand{\bibinfo}[2]{#2}
\providecommand{\BIBentrySTDinterwordspacing}{\spaceskip=0pt\relax}
\providecommand{\BIBentryALTinterwordstretchfactor}{4}
\providecommand{\BIBentryALTinterwordspacing}{\spaceskip=\fontdimen2\font plus
\BIBentryALTinterwordstretchfactor\fontdimen3\font minus
  \fontdimen4\font\relax}
\providecommand{\BIBforeignlanguage}[2]{{%
\expandafter\ifx\csname l@#1\endcsname\relax
\typeout{** WARNING: IEEEtran.bst: No hyphenation pattern has been}%
\typeout{** loaded for the language `#1'. Using the pattern for}%
\typeout{** the default language instead.}%
\else
\language=\csname l@#1\endcsname
\fi
#2}}
\providecommand{\BIBdecl}{\relax}
\BIBdecl

\bibitem{hofstede1980culture}
G.~Hofstede, ``Culture and organizations,'' \emph{International Studies of
  Management \& Organization}, pp. 15--41, 1980.

\bibitem{lenski2005ecological}
G.~E. Lenski, \emph{Ecological-evolutionary theory: Principles and
  applications}.\hskip 1em plus 0.5em minus 0.4em\relax Paradigm Publishers
  Boulder, CO, 2005.

\bibitem{landes1998wealth}
D.~S. Landes, ``The wealth and poverty of nations,'' \emph{WORLD AND I},
  vol.~13, pp. 258--263, 1998.

\bibitem{landes2000culture}
D.~Landes, ``Culture makes almost all the difference,'' \emph{Culture matters:
  how values shape human progress}, pp. 2--13, 2000.

\bibitem{gorodnichenko2011dimensions}
Y.~Gorodnichenko and G.~Roland, ``Which dimensions of culture matter for
  long-run growth?'' \emph{The American Economic Review}, vol. 101, no.~3, pp.
  492--498, 2011.

\bibitem{greif1994cultural}
A.~Greif, ``Cultural beliefs and the organization of society: A historical and
  theoretical reflection on collectivist and individualist societies,''
  \emph{Journal of political economy}, pp. 912--950, 1994.

\bibitem{triandis1985allocentric}
H.~C. Triandis, K.~Leung, M.~J. Villareal, and F.~I. Clack, ``Allocentric
  versus idiocentric tendencies: Convergent and discriminant validation,''
  \emph{Journal of Research in personality}, vol.~19, no.~4, pp. 395--415,
  1985.

\bibitem{gorodnichenko2010culture}
Y.~Gorodnichenko and G.~Roland, ``Culture, institutions and the wealth of
  nations,'' National Bureau of Economic Research, Tech. Rep., 2010.

\bibitem{singelis1995horizontal}
T.~M. Singelis, H.~C. Triandis, D.~P. Bhawuk, and M.~J. Gelfand, ``Horizontal
  and vertical dimensions of individualism and collectivism: A theoretical and
  measurement refinement,'' \emph{Cross-cultural research}, vol.~29, no.~3, pp.
  240--275, 1995.

\bibitem{world2012world}
W.~B. Group, \emph{World Development Indicators 2012}.\hskip 1em plus 0.5em
  minus 0.4em\relax World Bank Publications, 2012.

\bibitem{kopczuk2010earnings}
W.~Kopczuk, E.~Saez, and J.~Song, ``Earnings inequality and mobility in the
  united states: evidence from social security data since 1937,'' \emph{The
  Quarterly Journal of Economics}, vol. 125, no.~1, pp. 91--128, 2010.

\bibitem{wilkinson1997socioeconomic}
R.~G. Wilkinson, ``Socioeconomic determinants of health. health inequalities:
  relative or absolute material standards?'' \emph{BMJ: British Medical
  Journal}, vol. 314, no. 7080, p. 591, 1997.

\bibitem{acemoglu2002reversal}
D.~Acemoglu, S.~Johnson, and J.~A. Robinson, ``Reversal of fortune: Geography
  and institutions in the making of the modern world income distribution,''
  \emph{Quarterly journal of economics}, pp. 1231--1294, 2002.

\bibitem{acemoglu2012nations}
D.~Acemoglu, J.~A. Robinson, and D.~Woren, \emph{Why nations fail: the origins
  of power, prosperity, and poverty}.\hskip 1em plus 0.5em minus 0.4em\relax
  SciELO Chile, 2012, vol.~4.

\bibitem{diamond2005guns}
J.~M. Diamond and D.~Ordunio, \emph{Guns, germs, and steel}.\hskip 1em plus
  0.5em minus 0.4em\relax National Geographic, 2005.

\bibitem{brown2001cooperation}
S.~P. Brown and R.~A. Johnstone, ``Cooperation in the dark: signalling and
  collective action in quorum-sensing bacteria,'' \emph{Proceedings of the
  Royal Society of London. Series B: Biological Sciences}, vol. 268, no. 1470,
  pp. 961--965, 2001.

\bibitem{triandis1995individualism}
H.~C. Triandis, \emph{Individualism \& collectivism.}\hskip 1em plus 0.5em
  minus 0.4em\relax Westview Press, 1995.

\bibitem{razin1995population}
A.~Razin and E.~Sadka, \emph{Population economics}.\hskip 1em plus 0.5em minus
  0.4em\relax MIT Press, 1995.

\end{thebibliography}

\end{document}